# Chapter 2
# Addressing the Challenges in Federating Edge Resources

*Cihat Baktir[1], Cagatay Sonmez[1], Cem Ersoy[1], Atay Ozgovde[2], and Blesson Varghese[3]*

[1]*Computer Networks Research Laboratory, Department of Computer Engineering,*
*Bogazici University, Istanbul, Turkey*
[2]*Department of Computer Engineering, Galatasaray University, Istanbul, Turkey*
[3]*School of Electronics, Electrical Engineering and Computer Science, Queen's University Belfast, N. Ireland, UK*

## 2.1 Introduction

Edge computing is rapidly evolving to alleviate latency, bandwidth and Quality-of-Service concerns of Cloud-based applications as billions of 'things' are integrated to the Internet [1]. Current research has primarily focused on decentralizing resources away from centralized Cloud data centers to the Edge of the network and making use of them for improving application performance. Typically, Edge resources are configured in an ad hoc manner and an application or a collection of applications may privately make use of them. These resources are not publicly available, for example like Cloud resources. Additionally, Edge resources are not evenly distributed, but are sporadic in their geographic distribution.

However, ad hoc, private and sporadic Edge deployments are less useful in transforming the global Internet. The benefits of using the Edge should be equally accessible to both the developing and developed world for ensuring computational fairness and for connecting billions of devices to the Internet. However, there is minimal discourse on how Edge deployments can be brought to bear in a global context - federating them across multiple geographic regions to create a global Edge-based fabric that decentralizes data center computation. This of course is currently impractical, not only because of technical challenges, but is also shrouded by social, legal and geopolitical issues. In this chapter, we discuss two key challenges - networking and management in federating Edge deployments as shown in Figure 2.1. Additionally, we consider resource and modeling challenges that will need to be addressed for a federated Edge.

The key question we will be asking for addressing the networking challenge is "How can we create a dynamic enough networking environment that is compatible with the foreseen Edge computing scenarios in a federated setting" [2]. This is already a difficult issue for standalone and/or small-scale Edge deployments and requires further consideration in a federated setting. The dynamicity required is provided by the programmability of networking resources available through Software Defined Networking (SDN) in the today's context [3][4]. SDN with its northbound programming interface is an ideal candidate for the orchestration of Edge computing resources [5]. In the federated Edge context, however, a global coordination within SDN administrative domains would be needed. A harmony between local Edge deployments and the federated infrastructure is critical since both views of the system will be based on the same networking resources, possibly from competing perspectives. This will likely require a complete rethinking of the networking model for the Edge and further effort on the East-West interface of SDN. This chapter will discuss the networking challenges and will provide directions to resolve them.

As in any large-scale computing infrastructure, addressing management challenges becomes pivotal in offering seamless services. Currently, Edge-based deployments assume that services running on an Edge node either can be cloned or will be available on an alternate Edge node [6]. While this is a reasonable assumption for developing research in its infancy, it becomes a key challenge when federating Edge resources. In this context, future Internet architectures will need to consider how services can be rapidly migrated from one node to another based on demand [7]. Current technologies have limited scope in realizing this because of the large overheads and the lack of suitability for resource-constrained environments, such as the Edge. We will provide a discussion on management issues - benchmarking, provisioning, discovery, scaling and migration and provide research directions to address these [8][9].

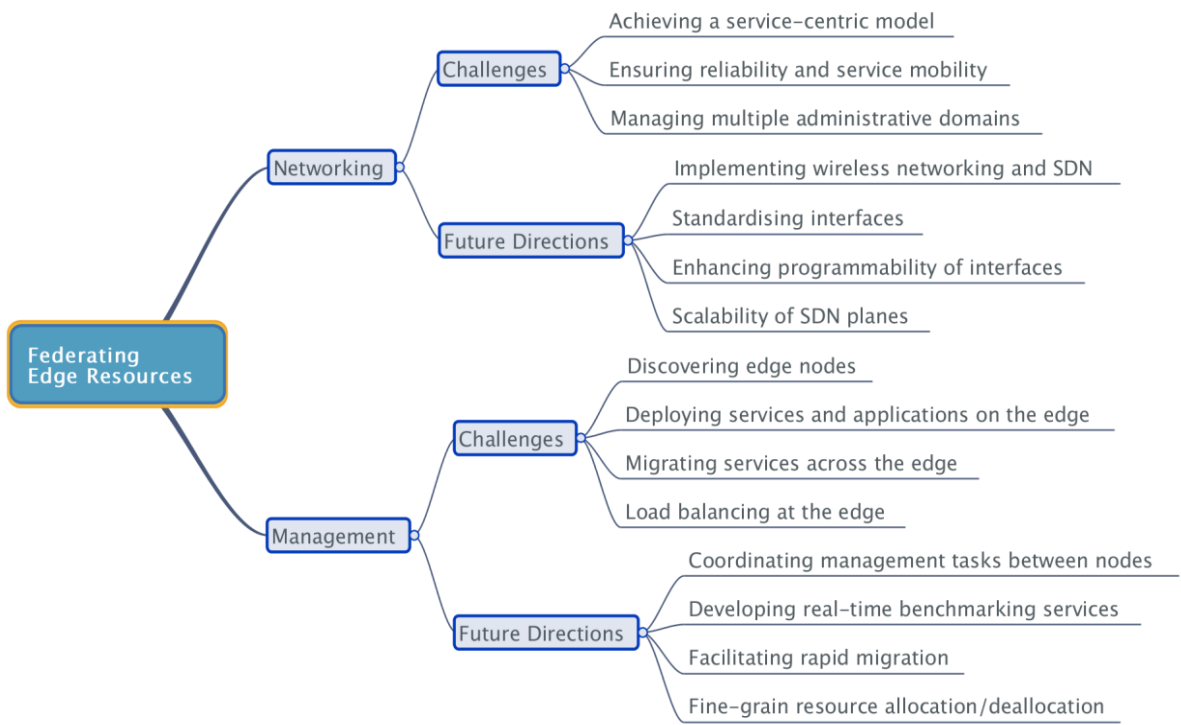

**Figure 2.1. Networking and management challenges in federating Edge resources**

Additionally, we present the resource and modeling challenges in this area. The resource challenge is related to both the hardware and software levels of resources that are employed at the Edge [10][11][12][13]. Although there are a number of reference architectures available for Edge based systems, we are yet to see a practical implementation of these systems. Often the hardware solutions are bespoke to specific applications and bring about heterogeneity at a significant scale that is hard to bridge. On the other hand, the software solutions that offer abstraction for making Edge resources publicly available have large overheads since they were designed for Cloud data centers.

The final consideration is the Modeling challenge. Integrating the Edge into the Cloud computing ecosystem brings about a radical change in the Internet architecture. It would be practically impossible to investigate and know the implications of large-scale Edge deployments, both from a technological and socio-economic perspective. A number of these can be modeled in simulators that offer the advantage of repeatability of experiments, minimizing hardware costs on experimental testbeds and testing in controlled environments [14]. However, our current understanding of interactions among users, Edge nodes and the Cloud is limited.

Throughout this chapter, we use the general term "Edge" to refer to a collection of technologies that aim towards decentralizing data center resources for bringing computational resources closer to the end user. Mobile Cloud Computing (MCC), Cloudlets, Fog computing, Multi-access Edge Computing (MEC) can all be considered as instances of Edge computing [9]. Therefore, generally speaking, the principles discussed in this chapter can be broadly applied to the aforementioned technologies.

The remainder of this chapter is organized in line with the above discussion and is as follows. Section 2.2 considers the networking challenge. Section 2.3 considers the management challenge. Section 2.4 presents the resource and modeling challenges. Section 2.5 concludes this chapter.

## 2.2 The Network Challenge

The network environment in which Edge servers will facilitate distributed computing is likely to be dynamic. This is because of the constantly varying demands at the end user level. The network infrastructure will need to ensure that the Quality-of-Service (QoS) of applications and services that will be deployed are not affected [15]. For this the quality of user experience cannot be compromised and the coordination of activities to facilitate Edge computing must be seamless and hidden from the end-user [16].

Table 2.1 summarizes the network challenges we consider in this chapter. The general networking challenge is in coping with the highly dynamic environment that the Edge is anticipated to be. This directly affects for example user mobility. As computational resources are placed closer to the source of the traffic, services become contextual. This results in the need for handling application layer handovers from one Edge node to another [17]. Depending on where the users are located and how request patterns are formed, the location of a service may change at any time. Another challenge is related to maintaining QoS in a dynamically changing environment.

| Networking Challenge | Why does it occur? | What is required? |
| --- | --- | --- |
| User mobility | Keeping track of different mobility patterns | Mechanisms for application layer handover |
| QoS in a dynamic environment | Latency-intolerant services, dynamic state of the network | Reactive behavior of the network |
| Achieving a service-centric model | Enormous number of services with replications | Network mechanisms focusing on "what" instead of "where" |
| Ensuring reliability and service mobility | Devices and nodes joining the network (or leaving) | Frequent topology update, monitoring the servers and services |

| Managing multiple administrative domains | Heterogeneity, separate internal operations and characteristics, different service providers | Logically centralized, physically distributed control plane, vendor independency, global synchronization |

**Table 2.1: Network challenges, their causes and potential solutions in federating Edge resources**

### 2.2.1 Network Challenges in a Federated Edge Environment

Federating Edge resources brings about a larger number of networking challenges related to scalability. For example, global synchronization between different administrative domains will need to be maintained in a federated Edge. Individual Edge deployments will have different characteristics, such as the number of services hosted and end-users in its coverage. In a federated context, different service offloads from multiple domains will need to be possible and will require synchronization across the federated deployments. In this section, we consider three challenges that will need to be addressed. We assume that Edge computing applications will be shaped by novel traffic characteristics that will leverage Edge resources possibly from different service providers.

The first challenge is in *achieving a service-centric model* on the Edge. The traditional host-centric model follows the 'server in a given geographic location' model, which is restrictive in a number of ways. For example, simply transferring a Virtual Machine (VM) image from one location to another can be difficult. However, in global Edge deployments, the focus needs to be on 'what' rather than 'where' so that services can be requested without prior knowledge of its geographic location [18]. In this model, services may have a unique identifier, may be replicated in multiple regions, and may be coordinated. However, this is not a trivial task given the current design of the Internet and protocol stacks, which do not facilitate global coordination of services.

The second challenge is in *ensuring reliability and service mobility*. User devices and Edge nodes may connect and disconnect from the Internet instantly. This could potentially result in an unreliable environment. A casual end-user device will be expecting seamless service perhaps via a plug and play functionality to obtain services from the Edge, but an unreliable network could result in latencies. The challenge here will be to mitigate this and create a reliable environment that supports the Edge. One mechanism to implement reliability is by either replicating services or by facilitating migration (considered in the management challenge) of services from one node to another. The key challenge here is to keep the overheads to a minimum so that the QoS of an application is not affected in any way.

The third challenge is in *managing multiple administrative domains*. The network infrastructure will need to be able to keep track of recent status of the network, Edge servers and services deployed over them. When a collection of end-user devices requires a service at the Edge, firstly the potential Edge host will need to be determined. The most feasible Edge node will then be chosen as the resource for the execution. There are two alternate scenarios that need to be considered for this operation: (i) the server is nearest to the end-users, or (ii) the potential server resides in another geographic region. Independent of the scenario, the network should forward the request to the server, and return the response to the end-user. During this progress, the data packets may travel across several distinct domains with multiple transport technologies. The challenge here is given this heterogeneity, user experience must not be compromised and the technical details may need to be concealed from the user device.

Addressing the above challenges require a solution that inherits the characteristics of both a centralized and distributed system. In order to achieve a global view of the network and maintain synchronization across separate administrative domains, the network orchestrator will need to follow a centralized structure. However, the control operations for coordinating the internal operations of a private domain will need to be distributed. In other words, the control of the network should be distributed over the network but should be placed within a logically centralized context.

### 2.2.2 Addressing the Network Challenge

We propose Software Defined Networking (SDN) as a solution for addressing the networking challenges as it naturally lends itself to handling them [5]. The key concept of SDN is to separate the control plane from the data plane and concentrate the core logic on a software-based controller [19]. The controller maintains the general view of the underlying network resources through its logically centralized structure [20]. This simplifies the management of the network, enhances the capabilities of the resources, and lowers the complexity barriers by utilizing resources more efficiently [21][22]. Most importantly, SDN facilitates instantaneous decision making in a dynamic environment by monitoring the status of the network at any time.

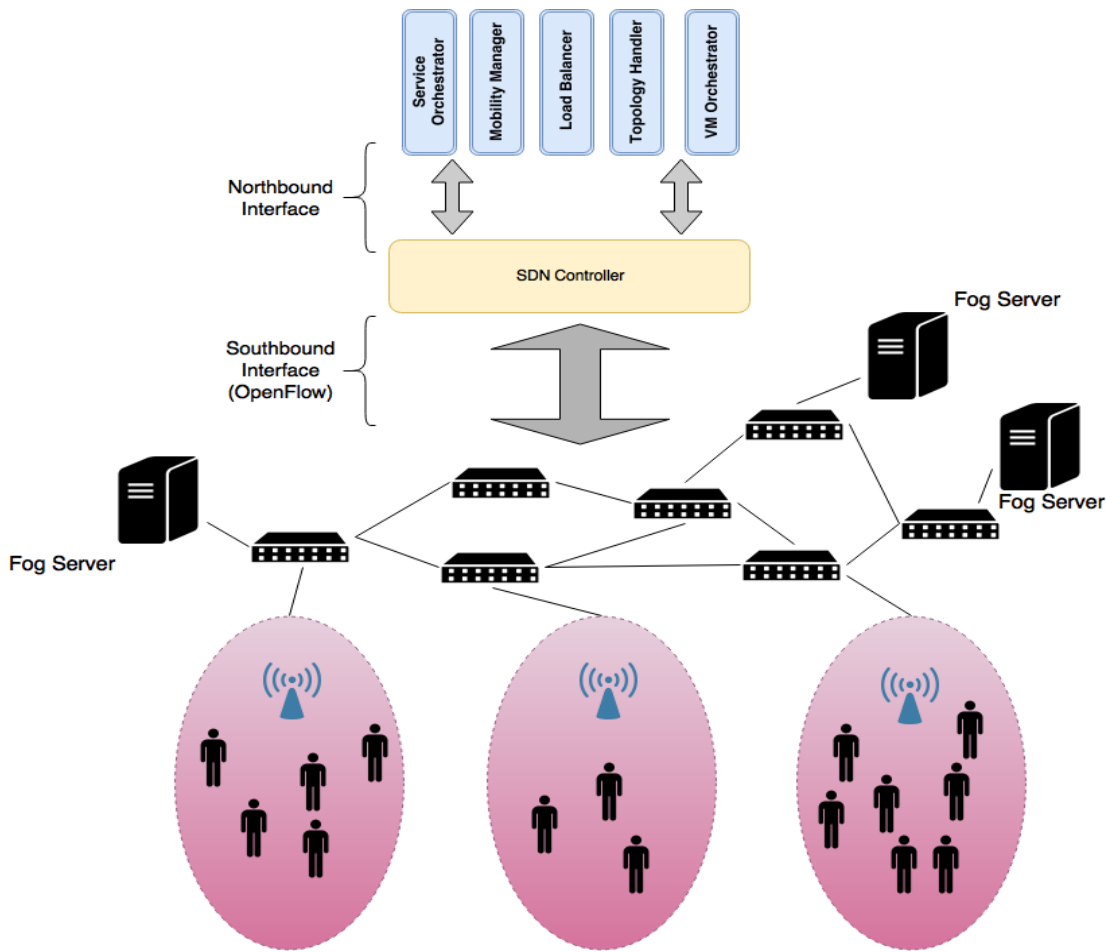

**Figure 2.2. Fog Computing with SDN as the Network Orchestrator**

The control plane communicates with the underlying network nodes through the OpenFlow protocol [23], which is considered as de-facto standard for the southbound interface of the SDN. On the other hand, the applications that define the behavior of the network communicate with the controller through the northbound interface, although it still remains to be standardized [24][25].

The programmable control plane can be either centralized or distributed physically. The initial proposal of SDN and OpenFlow considers a campus environment, and the design criteria were based on a single controller assuming that the control channel can handle the typical area of coverage. However, the novel Edge computing scenarios demand more than this. In order to make Edge deployments publicly accessible and to construct a global pool of computing resources at the Edge, the control plane should be distributed to enable the orchestration with multiple control instances. A typical SDN-orchestrated Edge computing environment is depicted in Figure 2.2 where the network devices are aligned with the SDN controller and related northbound applications.

The logically centralized scheme of the control plane is a key feature to manage user mobility by simplifying the management of the connected devices and resources [26]. When a new device is connected to the network or authenticated to another network due to mobility, the network should react as soon as possible and provides the plug and play functionality. This capability is granted to the SDN controller with a functionality of topology discovery through the OpenFlow Discovery (OFDP) protocol [27]. As soon as the state of an end-user changes, the corresponding flow rules are immediately updated by the controller. Through a module implemented as a northbound application, the topology is checked frequently and any newly added, disconnected or modified node can be updated on the topology view. The opportunity here is that the node can be either an end-user device, a computational resource or a switch. The controller can handle the integration of each type of component while updating the topology view. This approach also enables to handle the application layer handover, which is triggered by the mobility during service offloading.

The utilization of OpenFlow-based switches, such as OpenvSwitch [28], and SDN controller as the umbrella of the whole system enhances the effectiveness of the control through a better resolution of management [29]. Considering the northbound applications that define the behavior of the network through user-defined policies, the network could be reactive or proactive. For instance, in a university campus environment, students and university staff are always in motion. The traffic flow increases during daytime and decreases after working hours. In this single administrative domain, where mobility is high, the reactive operations come into prominence. Through the ability of gathering statistical information, such as the traffic load forwarded by a certain node or link, from the data plane elements by exchanging OpenFlow messages, the SDN controller may dictate flow rules that lead to near optimal solutions within the network. In the case of Edge computing, where multi-tenants share resources and application instances enforce strict QoS criteria in terms of latency, the SDN controller can modify the flow rules at the Edge of the network if an Edge server becomes highly loaded or a probability of congestion emerges. The SDN controller not only monitors the status of the networking nodes and links, it can also be integrated with a server monitor functionality through a northbound application. Therefore, one can define a customized policy that is able to provide a load-balancing algorithm considering both computing and networking resources.

Federating Edge resources creates a globally accessible infrastructure, but makes the environment more dynamic. Managing mobility within a single domain, handling the application layer handover among the servers in the same vicinity and reacting to the changes due to a set of users can be leveraged by a single control plane component. However, the realistic and practical approaches of Edge computing deployments necessitate a network behavior to support flexibly the operations of a federated setting in a global context. As might be expected, a single control plane cannot satisfy the global management of various types of devices and administrative domains. The evolution of SDN and OpenFlow allows for a logically centralized but physically distributed control plane. The data traffic may be forwarded through at least two different domains that belong to separate service providers. Therefore, there is a need for abstraction and control over the disjoint domains with multiple controllers.

A controller can be deployed for handling the operations within a single domain. However, there will be the need for inter-domain or inter-controller communication to maintain reliability in forwarding traffic to the gateways. This communication is provided by the east-west interface. The control plane can be organized as either hierarchical or flat. In the hierarchical structure, a master plane provides the synchronization among the domains. The lower-level controllers are responsible for their own domains. If an event occurs within a domain, the corresponding controller can update the other controllers by informing the master controller. When a flat structure is utilized, the controllers communicate directly with each other to achieve synchronization through their east-west interfaces.

In a federated Edge setting, the distributed control plane will play an important role for addressing scalability and consistency issues. Considering a service-centric environment, multiple controllers should simultaneously handle coordinating the service replications and tracking their locations. Without the flexibility provided by SDN and programmable networks [30], it requires extra effort to implement the service-centric design. Since SDN can intrinsically retrieve a recent view of the underlying network, controllers can keep track of the locations of the services. A northbound application that maps the service identifiers to the list of locations paves the road for embedding the service-centric model into the global Edge setting. Whenever a user offloads a service by specifying its identifier, the controller responsible for that domain can inspect this information and retrieve the list of possible destinations. The list of possible destinations is frequently updated by communicating with the remaining controllers that are coordinating other administrative domains. With the help of an adjacent load balancing northbound application, the network can determine the most feasible server and forward the request by modifying the header fields of the packet. If the destination is deployed in another region, the forwarding operation requires the packets to be routed over multiple domains and it is handled by the cooperative work of the distributed control plane. If a new service is deployed within a region or a replication of a service is created, the responsible controller initially updates its database and creates an event to inform the other controllers to keep them synchronized. The OpenFlow messages exchanged through the east-west interfaces provide the global synchronization in case of an event.

Service mobility needs to be addressed in the context of a federated Edge. Creating, migrating and replicating services need to be accomplished at the Edge to deal with varying traffic patterns and load balancing [31]. SDN again is a candidate solution since it determines possible destination nodes and the path that can be taken for migrating a service such that the performance is least affected (congestion is prevented) [32]. The operation in SDN can be carried out using flow rules.

In research and experiments, SDN is known for managing and handling heterogeneity well [33][34]. From a network perspective, federated Edge environments will typically be heterogeneous in that they will comprise different network types in addition to the varying traffic flow patterns. In this case, the control plane could provide an interoperable-networked environment comprising multiple domains that belong to different providers for both Edge servers and end-user devices. Additionally, vendor dependency and compatibility issues between different networking devices are eliminated [35].

### 2.2.3 Future Research Directions

The integration of Fog computing and SDN has immense potential for accelerating practical deployments and federating resources at the Edge. However, there are avenues that still need to be explored for bridging the gap between Fog computing and SDN. In this section, we consider four such avenues as directions for future research:

1. **The implementation of wireless networking and SDN.** Existing research and practical implementations achieve network virtualization via SDN. However, the focus is usually on the virtualization and management of SDN controllers in a wired network [36]. We believe to exploit the benefits of SDN and current standards, such as OpenFlow, fully also for wireless networks for federating Edge nodes which will be serving mostly a mobile community in the future.
2. **Standardization of interfaces for interoperability.** OpenFlow is currently the de facto standard for the southbound interface; however, there are no recognized standards for northbound communication (although the Open Networking Foundation (ONF) for northbound standards [37] organizes a working group). Lack of standardization prevents interoperability among northbound applications that run on top of the same controller. We believe that developing standards for the northbound applications is another important avenue for future research. Additionally, existing SDN-based scenarios do not depend on the east-west interface and there is very little research in this area. Communication between the adjacent controllers needs to become more reliable and efficient in order not to burden the control channels. We believe that focusing efforts towards this area will provide opportunities for federating a pool of computational resources at the Edge.
3. **Enhancing programmability of existing standards and interfaces.** Our experience in programming using OpenFlow leads us to recommend the implementation of additional functionalities. The recent version of OpenFlow (v1.5.1) provides only partial programmability within the network. In order to enable federation of Edge resources for generic Fog deployments, we believe additional research is required for enhancing the programmability of standards and interfaces.
4. **Scalability of SDN planes for reaching wider geographic areas.** It is anticipated that Edge nodes will be distributed over a wide geographic area and there will be distinct administrative domains in Fog computing systems. Here, a distributed form of the SDN control plane is required to communicate with the adjacent controllers. Therefore, we recommend further investigation of the scalability of SDN planes [38]. This is challenging because there are no standards in place for east-west interfaces, which will need to be utilized for the inter-controller communication.

## 2.3 The Management Challenge

Adding a single layer of Edge nodes in between the Cloud and user devices introduces significant management overheads. This becomes even more challenging when clusters of Edge nodes need to be federated from different geographic locations to create a global architecture. In this section, we consider four management challenges that will need to be addressed and are presented in Table 2.2.

The first management challenge is related to *discovering Edge resources* both at individual and collective levels. At the individual level, potential Edge nodes that can provide computing will need to be visible in the network, both to applications running on user devices and their respective cloud servers. At the collective levels, a collection of Edge nodes in a given geographical location (or at any other granularity) will need to be visible to another collection of Edge nodes. In addition to the system challenges, assuming that an Edge node has near similar capabilities of a network device and general purpose computational device, the challenge here is to determine the best practice for discovery - whether discovery of Edge nodes is (i) self-initiated and results in a loosely coupled collection, (ii) initiated by an external monitor that results in a tightly coupled collection, or (iii) a combination of the former.

| Management challenge | Why should it be addressed? | What is required? |
|---|---|---|
| Discovery of Edge nodes | To select resources when they are geographically spread and loosely coupled | Light-weight protocols and handshaking |
| Deployment of service and applications | Provide isolation for multiple services and applications | Monitoring and benchmarking mechanisms in real-time |
| Migrating services | User mobility, workload balancing | Low overhead virtualization |
| Load balancing | To avoid heavy subscription on individual nodes | Auto-scaling mechanisms |

Table 2.2: Management challenges, the need for addressing them and potential solutions in federating Edge resources

The second management challenge is related to *deploying services and applications on the Edge*. Typically, a service that can furnish requests from user devices will need to be offloaded onto one or a collection of Edge nodes. However, this will not be possible without knowing the capabilities of the target Edge and matching them against the requirements of services or applications (such as the expected load and amount of resources required), given that there may be multiple clusters of Edge nodes available in the same geographic location. Benchmarking multiple Edge nodes (or multiple collections) simultaneously will be essential here to meet the service objectives. This is challenging and will need to be performed in real-time.

The third management challenge is related to *migrating services across the Edge*. Existing technology allows for deploying applications and services using Virtual Machines (VMs), containers and unikernel technologies. These technologies have proven to be useful in the Cloud context to deploy an application and migrate them across data centers. Given the availability of significant resources in a Cloud data center it is not challenging to maintain a large repository of images that can be used to start up or replicate services in the event of failures or load balancing. This however is challenging on the Edge given the real-time and resource constraints. Additionally, the shortest path in the network for migrating services from an Edge node to another will need to be considered.

The fourth management challenge is related to the **load balancing at the Edge**. If there is significant subscription of services at the Edge, then the resource allocation for individual services on a single Edge node or in the collection will need to be managed. For example, if there is one service that is heavily subscribed when compared to other services that are dormant on the Edge, then the resources allocated to the heavily subscribed service will need to be scaled. While this is just one scenario, it becomes more complex when more services require resources from the same collection of Edge nodes. This will require significant monitoring of resources at the Edge, but traditional methods cannot be employed given the resource constraints on Edge nodes. Similarly, mechanisms will need to be put in place for scaling the resources for one service (which may be heavily subscribed) while de-allocating resources from dormant services. Both the monitoring and scaling mechanisms will need to ensure integrity so that the workload is fairly balanced.

### 2.3.1 Current Research

Existing techniques for discovery of Edge nodes can be classified based on whether they operate in a multi-tenant (i.e. more than one service can be hosted on the Edge node) environment or not. For example, FocusStack discovers Edge nodes in a single tenant environment [39], whereas ParaDrop [40] and Edge-as-a-Service (EaaS) [41] operate in multi-tenant Edge environments. However, there are additional challenges that will need to be addressed to enable discovery when multiple collections of Edge nodes are federated.

Current research on deploying services focus on pre-deployment resource provisioning (matching requirements of an application against available resources before the application is deployed) [42]. Post-deployment becomes even more important both in the context of individual Edge nodes and federated Edge resources due to variability (more applications need to be hosted on a collection of Edge nodes) of workloads that are anticipated on the Edge. Additionally, workload deployment services that operate on distributed clusters focus on large jobs, such as Hadoop or MapReduce [43][44]. However, post-deployment techniques suitable for more fine-grained workloads will be required for federated Edge resources.

Migration of services via Virtual Machines (VMs) across clusters is possible, but in reality has a significant time overhead [45][46]. Additionally, live migration across geographically distributed Cloud data centers is more challenging and time consuming. Similar strategies have been adopted in the context of Edge resources for live migration of VMs [47][48]. While this is possible although migration takes a few minutes it is still challenging to use existing strategies for real-time use. Additionally, VMs may not be the de facto standard for hosting services on the Edge [11][49]. Alternate lightweight technologies, such as containers, and how they may be used to migrate workloads at the Edge will need to be investigated and the strategies underpinning these will need to be incorporated within container technologies.

Monitoring of Edge resources will be a key requirement for achieving load balancing. For example, performance metrics will need to be monitored for implementing auto-scaling methods to balance workloads on the Edge. Existing monitoring systems for distributed systems either do not scale or are resource consuming. These are not suitable for large-scale resource constrained Edge deployments. Current mechanisms for auto scaling resources are limited to single Edge nodes and employ lightweight monitoring [11]. However, scaling these mechanisms are challenging.

### 2.3.2 Addressing the Management Challenges

Three of the above four research challenges, namely discovery, deployment and load balancing were addressed in the context of individual Edge nodes at Belfast on the Edge-as-a-Service platform and the ENORM framework.

**Edge-as-a-Service (EaaS) Platform**

The Edge-as-a-Service (EaaS) [41] platform targets the discovery challenge and implements a lightweight discovery protocol for a collection of homogeneous Edge resources (Raspberry Pis). The EaaS platform operates in a three-tier environment - top tier is the Cloud, bottom tier comprises user devices and middle tier contains Edge nodes. The platform requires a master node, which may either be a compute available network device or a dedicated node and executes a manager process that communicates with the Edge nodes. The master node manager communicates with potential Edge nodes and installs a manager on the Edge node to execute commands. Administrative control panels are available on the master node to monitor individual Edge nodes. Once the EaaS platform discovers an Edge node, then Docker or LXD containers can be deployed. The platform was tested in the context of an online game, similar to the popular PokeMon Go, to improve the overall performance of the application.

The benefit of this platform is that the discovery protocol that is implemented is lightweight and the overhead is a few seconds for launching, starting, stopping or terminating containers. Up to 50 containers with the online game workload were launched on an individual Edge node. However, this has been carried out in the context of a single collection of Edge nodes. Further research will be required to employ such a model in a federated Edge environment.

The major drawback of the EaaS platform is that it assumes a centralized master node that can communicate with all potential Edge nodes. The research also assumes that the Edge nodes can be queried and can via owners be made available in a common marketplace. Additionally, the security related implications of the master node installing a manager on the Edge node and executing commands on the Edge node is not considered.

**Edge NOde Resource Management (ENORM) Framework**

The ENORM framework [11] primarily addresses the deployment and load balancing challenges on individual Edge nodes. Similar to the EaaS platform ENORM operates in a three-tier environment, but a master controller does not control the Edge nodes, instead it is assumed that they are visible to Cloud servers that may want to make use of the Edge. The framework allows for partitioning a Cloud server and offloading it to Edge nodes for improving the overall Quality-of-Service (QoS) of the application.

The framework is underpinned by a provisioning mechanism for deploying workloads from a Cloud server onto an Edge server. The Cloud and an Edge server establish a connection via handshaking to ensure that there are sufficient resources available to fulfil the request of the server that will be offloaded onto the Edge. The provisioning mechanism caters for the entire lifecycle of an application server from offloading it onto the Edge via a container until it is terminated and the Cloud server is notified.

Load balancing on a single Edge node is accomplished by implementing an auto-scaling algorithm. It is assumed that an Edge node could be a traffic routing node, such as a router or mobile base station and therefore an offloaded service should not compromise the QoS of the basic service (traffic routing) that is executed on the node. Each application server executing on the Edge node has a priority. Each Edge server is monitored (in terms of both network and system performance) and it is estimated whether the QoS can be met. If an Edge server with a higher priority cannot meet its QoS, then the resources for the application is scaled. If the resource requirements of an application cannot be met on the Edge then it is moved back to the Cloud server that offloaded it. This occurs iteratively in periodic intervals to ensure that the QoS is achieved and the node is stable.

The ENORM framework is also validated on the online game use-case as for the EaaS platform. It is noted that the application latency can be reduced between 20% - 80% and the overall data transferred to the Cloud for this use-case is reduced by up to 95%.

### 2.3.3 Future Research Directions

Both the EaaS platform and the ENORM framework have limitations in that they do not assume federated Edge resources. In this section, the following four research directions for addressing management challenges when federating Edge resources are considered:

1. ***Coordinating management tasks between heterogeneous nodes of multiple Edge collections***. Federating Edge resources inevitably requires the bringing together heterogeneous Edge nodes (routers, base stations, switches and dedicated low power compute devices). While managing homogeneous resources in itself can be challenging, it will be more complex to coordinate multiple collections of heterogeneous resources. The challenge here is enabling the required coordination via a standard protocol to facilitate management between devices that are geographically apart, that have varying CPU architectures, and may inherently be used for network traffic routing.
2. ***Developing real-time benchmarking services for federated Edge resources***. Given the varying computational capabilities and workloads on (traffic through) Edge nodes, cloud servers will need to benchmark reliably a portfolio of Edge nodes. Either this portfolio may be from different or the same geographic location, so that via benchmarking the application server can identify Edge nodes that may meet Service Level Objectives (SLOs) if a partitioned workload needs to be deployed on the Edge. Mechanisms facilitating this in real-time will need to be developed.
3. ***Facilitating rapid migration between federated Edge resources***. Current migration techniques typically have overheads in the order of a few minutes in the best cases when attempting to migrate from one node to another. This overhead will obviously increase with geographic distance. Current mechanisms for migration take a snapshot of the VM or the container on an Edge node and then transport this across to another node. To facilitate fast migration perhaps alternate virtualization technologies may need to be developed that allow for migration of more abstract entities (such as functions or programs). This technology may also be employed in upcoming serverless computing platforms for developing interoperable platforms across federated Edge resources.

4. ***Investigating fine-grain resource allocation/deallocation for load balancing using auto-scaling***. Current auto-scaling methods add or remove discrete predefined units of resources on the Edge for auto-scaling. However, this is limiting in resource-constrained environments in that resources may be over-provisioned. Alternate mechanisms will need to be investigated that can derive the amount of resources that need to be allocated/deallocated based on specific application requirements to meet SLOs without compromising the stability of the edge environment.

## 2.4 Miscellaneous Challenges

The previous two sections have considered the networking and management challenges in federating Edge resources that are geographically distributed. However, there are additional challenges that need to be considered. For example, the challenge of developing pricing models to make use of Edge resources. This will rely on a solution space that cannot be fully foreseen today given that the technology for supporting public Edge computing is still in its infancy. In this section, we consider two further challenges, namely the resource and modeling challenges as shown in Figure 2.3, which are dependent on both networking and management.

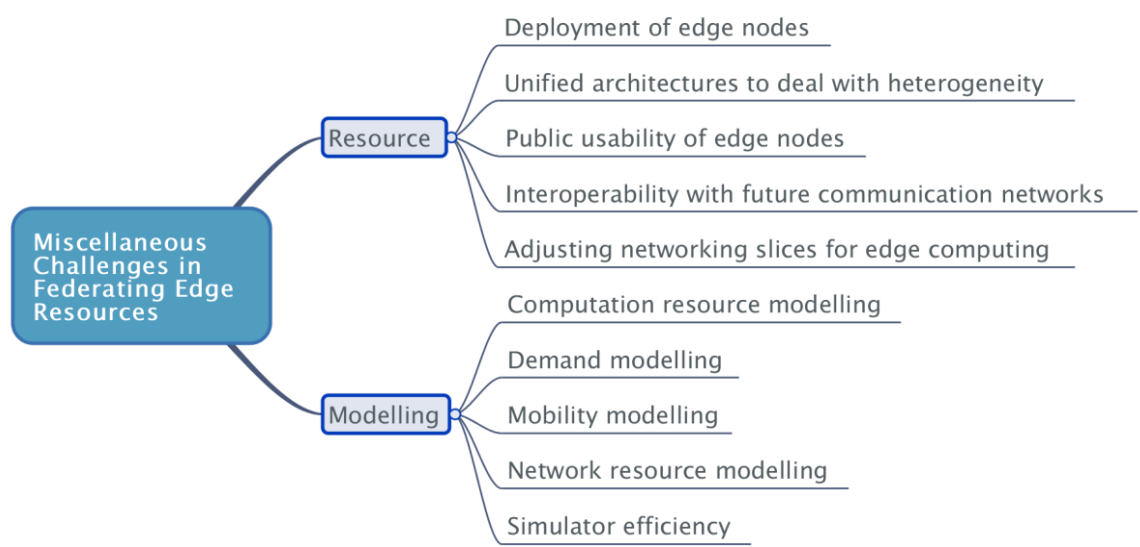

Figure 2.3. Resource and Modeling Challenges in federating Edge resources

### 2.4.1 The Research Challenge

The prospect of including an Edge layer between Cloud data centers and user devices for emerging applications is appealing because latency and transmission of data to the core of the network can be minimized to improve the overall Quality-of-Service of an application. Although there are reference Edge architectures and test beds that validate these architectures, Edge computing is not yet publicly adopted and we are yet to see large-scale practical implementations of these systems. We present five resource challenges that will need to be addressed before an Edge layer can become a practical reality.

The first resource challenge is related to **deploying Edge nodes**. It is still not clear whether Edge nodes are likely to be (i) traffic routing nodes, such as routers, switches, gateways and mobile base stations that integrate general purpose computing via CPUs on them, (ii) dedicated computing nodes with low power compute devices on which general purpose computing can be achieved, such as micro Clouds, or (iii) a hybrid of the former. In the retail market, products that enable general purpose computing on traffic routing nodes are available. For example, Internet gateways that are Edge enabled are currently available in the market[1]. Additionally, there is ongoing research that aims to use micro Cloud data centers at the Edge of the network[2]. It seems that there is a business use case for either option, but how the latter will coexist with traffic routing nodes is yet to be determined. Furthermore, the migration to the former may take a long time since existing traffic routing nodes will need to be upgraded.

The second resource challenge is related to **developing unified architectures to account for heterogeneity**. Bringing different types of Edge-based nodes with varying performance and compute resources as a coherent single layer or multiple layers can be challenging from a software, middleware and hardware perspective. Given the wide variety of Edge computing options proposed ranging from small home routers to micro Cloud installations, federating them will require the development of unified interoperable standards across all these nodes. This is unprecedented and will be unlike the standards that have been used on the Cloud where large collections of compute resources have the same underlying architecture. If this were the case, then applications and services will need to be executed in a manner oblivious to the underlying hardware. However, current research that enables this via virtualization or containerization is not suitable and is not available for all hardware architectures.

The third resource challenge is related to **public usability of Edge nodes**. Regardless of how the Edge layer is enabled, it is anticipated to be accessible both for bringing computation from Clouds closer to the user devices and for servicing requests from user devices or processing data generated from a large collection of sensors before it is sent to the Cloud. This raises several concerns - (i) How will Edge nodes be audited? (ii) Which interface will be used to make them publicly accessible? (iii) Which billing model will be required? (iv) Which security and privacy measures will need to be adopted on the Edge? These concerns are beyond the scope of this chapter. However, they will need to be addressed for obtaining publicly usable Edge nodes.

The fourth resource challenge is related to **interoperability with future communication networks**. In the context of Edge computing systems, the network itself is a critical resource that defines the overall performance of an Edge solution. Resource management strategies should consider the network resources as well as the computational resources for the efficient operation of the Edge systems [14]. Initial Edge computing proposals almost exhaustively employ WLAN technologies for accessing computational resources. However, this is likely to change given the emergence of 5G. QoS provided at the level of tactile Internet makes 5G systems a strong alternative for Edge access [52]. Considering the potential of Edge systems, European Telecommunications Standards Institute (ETSI) started the Multi-Access Edge Computing (MEC) standardization with many contributors from the telecom industry [53]. MEC, in principle, is an Edge computing architecture that is envisioned as an intrinsic component of 5G systems. With 5G practical deployments whether Edge computing services will rely on the MEC functionality or whether they will make use of high bandwidth 5G network capabilities and position themselves as an Over-The-Top (OTT) is not yet clear and will depend on parameters including cost and openness. The two options will dictate radically different positions for the federation of Edge computing systems. ETSI-MEC with its inherent position within the 5G architecture will closely couple Edge systems in general and their federation with the telecom operators' operation of the whole network.

---

[1] http://www.dell.com/uk/business/p/edge-gateway

[2] http://www.dell.com/en-us/work/learn/rack-scale-infrastructure

The fifth resource challenge is related to **adjusting network slices for Edge systems**. Another important opportunity that is anticipated to be provided by future networks is slicing. Network slices are defined as logical networks overlaid on a physical or virtual network that can be created on demand with a set of parameters [52]. Slicing will allow network operators to cater for QoS specific to a service or a group of services. Although, slicing is not an approach that is particular to Edge computing, it will significantly affect the operation and the performance of Edge systems. An end-to-end slice dedicated to each service on an Edge server would be beneficial. However, this becomes challenging as the slices become fine grained due to scalability and management challenges. In addition, slicing related to Edge computing needs to consider the volume of interaction between Edge and Cloud servers. A simple approach would be to assign slices for every Edge deployment and expect the Edge orchestration system to assign additional resources within the Edge system. In a federated setting, a finite capacity of resources may be assigned to a set of standalone Edge systems and the overall slice of the individual Edge system may be adjusted globally considering resource usage patterns.

### 2.4.2 The Modeling Challenge

Edge Computing has paved way for a variety of technologies, such as Mobile Cloud Computing (MCC), Cloudlets, Fog computing and Multi-access Edge Computing (MEC) [5]. These indicate that Edge solutions can be obtained in multiple domains using different techniques. Given that there are no de facto standards and that there is an abundance of Edge architectures emerging in the literature, tools for modeling and analyzing Edge systems are required.

One option to model Edge systems is to implement test beds that are specific to the requirements of a use case. Given the availability of open source tools to virtualize resources (computational and network), it would be feasible to develop test beds for a research environment. For example, the Living Edge Lab[3] is an experimental testbed. However, setting up testbeds can be quite expensive. Additionally, for a complete performance analysis, testbeds and sometimes even real world deployments may not lend themselves to repeatable and scalable experiments as could be obtained using simulators [50]. Therefore, simulators are employed to complement experimental testbeds for a thorough evaluation.

At the heart of a simulator is a complex mathematical model that captures the environment. Although simulators are advantageous, numerous modeling challenges will need to be addressed in designing an ideal (or even a reasonable) simulator [51]. An ideal simulation environment should incorporate programming APIs, management of configuration files and UI dashboards for easy modeling with minimum manual effort. We anticipate that the same principles would apply to an Edge simulator. In this section, we consider five specific modeling challenges that will need to be addressed for an ideal Edge simulator.

---

[3] http://openedgecomputing.org/lel.html

The first challenge is related to ***computational resource modeling***. Like a Cloud data center, an Edge server will provide computational power to its users via virtualization techniques, such as Virtual Machines and containers [12]. A simulation environment in this context should support the creation, resizing, migration of virtual resources and model CPU, memory and network resource consumption at different levels of granularity (process, application and entire node). The model will need to capture the possibilities of using existing traffic routing nodes, dedicated nodes or a combination of these.

The second challenge is related to ***demand modeling***. To be able model the load on an Edge computing system, the demand on the Edge resource due to an individual user (or a collection of users) will need to be modeled. Accounting for the heterogeneity of mobile devices and the traffic generated by a variety of applications is complex. End user devices or a Cloud server may offload compute on to Edge servers and this will need to be accounted for in a demand model. The distribution of demand and the inter-arrival times of traffic on the Edge will need to be considered. Profiles of the users and/or family of applications with predefined distributions would also be beneficial.

The third challenge is related to ***mobility modeling***. Mobility is a key component that will need to be considered for accurately modeling the time-varying demands on the Edge. The need for mobility arises in multiple use-cases. For example, a human with a wearable gadget moving from the coverage area of one Edge server to another could result in the migration of the service from one Edge server onto another or the replication of the service with the user data on another Edge server. In such a use case, the simulation environment should allow for designing experiments that realistically captures mobility in a variety of forms.

The fourth challenge is related to ***network modeling***. Performance and behavior patterns of the network are critical for the overall operation of an Edge system. Accurate network delay modeling will not be easy due to dynamic workloads that operate using different network access technologies, such as Wi-Fi, Bluetooth and cellular networks. In contrast to legacy network simulators, an Edge simulation tool should be able to scale rapidly network resources. This requirement arises due to the *slicing* approach described previously in which multiple network slices will need to be modeled in the network [52].

The final challenge considered is the ***simulator efficiency***. Simulators will need to be scalable, extensible to changing infrastructure requirements and easy to use. Taking into account the upcoming Internet-of-Things and Machine-to-Machine communications, the time complexity of simulators accounting for federated Edge resources should model the connections of a large number of devices and users.

## 2.5    Conclusion

Computational resources that are typically concentrated in Cloud data centers are now proposed to become available at the edge of the network via Edge computing architectures. Edge resources will be geographically distributed and they will need to be federated for a globally accessible Edge layer that can service both data center and user device requests. The aim of this chapter is to highlight some of the challenges that will need to be addressed for federating geographically distributed Edge resources. The chapter firstly presented the network and management related issues. Then the chapter considered how existing research reported in the literature addresses these challenges and provided a roadmap of future directions. Subsequently, we presented additional challenges related to resources and modeling for a federated Edge. The key message of this chapter is that federating Edge resources is not an easy task. Let alone the social and legal aspects in federating, underlying technologies that will facilitate public Edge computing is still in its infancy and rapidly changing. There are a number of networking, management, resource and modeling related technological challenges that will need to be addressed for developing novel solutions to make the federated Edge computing a reality.

**References**

[1] B. Varghese, and R. Buyya, Next generation Cloud computing: New trends and research directions, *Future Generation Computer Systems*, 79(3): 849-861 (February 2018).

[2] W. Shi, and S. Dustdar, The promise of Edge computing, *Computer*, 49(5): 78-81 (May 2016).

[3] T. Taleb, K. Samdanis, B. Mada, H. Flinck, S. Dutta, and D. Sabella, On multi-access Edge computing: A survey of the emerging 5G network Edge architecture and orchestration, *IEEE Communications Surveys & Tutorials*, 19(3): 1657-1681 (May 2017).

[4] R. Vilalta, A. Mayoral, D. Pubill, R. Casellas, R. Martínez, J. Serra, and R. Muñoz. End-to-End SDN orchestration of IoT services using an SDN/NFV-enabled Edge node, in *Proceedings of Optical Fiber Communications Conference and Exhibition*, Anaheim, CA, USA, March 20-24, 2016.

[5] A. C. Baktir, A. Ozgovde, and C. Ersoy, How can Edge computing benefit from software-defined networking: A survey, Use Cases & Future Directions, *IEEE Communications Surveys & Tutorials*, 19(4): 2359-2391 (June 2017).

[6] T. Q. Dinh, J. Tang, Q. D. La, and T. Q. S. Quek, Offloading in Mobile Edge computing: Task allocation and computational frequency scaling, *IEEE Transactions on Communications*, 65(8): 3571-3584 (August, 2017).

[7] L. F. Bittencourt, M. M. Lopes, I. Petri, and O. F. Rana, Towards virtual machine migration in Fog computing, *10th International Conference on P2P, Parallel, Grid, Cloud and Internet Computing*, Krakow, Poland, November 4-6, 2015.

[8] J. Xu, L. Chen, and S. Ren, Online learning for offloading and autoscaling in energy harvesting Mobile Edge computing, *IEEE Transactions on Cognitive Communications and Networking*, 3(3): 361-373 (September 2017).

[9] N. Apolónia, F. Freitag, L. Navarro, S. Girdzijauskas, and V. Vlassov, Gossip-based service monitoring platform for wireless Edge Cloud computing, in *Proceedings of the 14th International Conference on Networking, Sensing and Control*, Calabria, Italy, May 16-18, 2017.

[10] M. Satyanarayanan, Edge computing: Vision and challenges, *IEEE Internet of Things Journal*, 3(5): 637-646 (June 2016)

[11] N. Wang, B. Varghese, M. Matthaiou, and D. S. Nikolopoulos, ENORM: A framework for Edge node resource management, *IEEE Transactions on Services Computing*, PP(99): 1-1 (September 2017).

[12] B. Varghese, N. Wang, S. Barbhuiya, P. Kilpatrick, and D. S. Nikolopoulos, Challenges and opportunities in Edge computing, in *Proceedings of the International Conference on Smart Cloud*, New York, USA, November 18-20, 2016.

[13] Z. Hao, E. Novak, S. Yi, and Q. Li, Challenges and software architecture for Fog computing, *IEEE Internet Computing*, 21(2): 44-53 (March 201).
[14] C. Sonmez, A. Ozgovde, and C. Ersoy, EdgeCloudSim: An environment for performance evaluation of Edge computing systems, in *Proceedings of the 2nd International Conference on Fog and Mobile Edge Computing*, Valencia, Spain, May 8-11, 2017.
[15] S. Yi, C. Li, and Q. Li, A survey of Fog computing: Concepts, applications and issues. in *Proceedings of the Workshop on Mobile Big Data*, Hangzhou, China, June 22-25, 2015.
[16] L. M. Vaquero, and L. Rodero-Merino, Finding your way in the Fog: Towards a comprehensive definition of Fog computing, *SIGCOMM Computer Communication Review*, 44(5): 27-32 (October 2014).
[17] I. Stojmenovic, S. Wen, X. Huang, and H. Luan, An overview of fog computing and its security issues, *Concurrency and Computation: Practice and Experience*, 28(10): 2991-3005 (April 2015).
[18] A. C. Baktir, A. Ozgovde, and C. Ersoy, Enabling service-centric networks for Cloudlets using SDN. in *Proceedings of the 15th International Symposium on Integrated Network and Service Management*, Lisbon, Portugal, May 8-12, 2017.
[19] H. Farhady, H. Lee, and A. Nakao, Software-Defined Networking: A survey, *Computer Networks*, 81(C): 79-95 (December 2014).
[20] R. Jain, and S. Paul, Network virtualization and software defined networking for Cloud computing: A survey, *IEEE Communications Magazine*, 51(11): 24-31, 2013.
[21] M. Jammal, T. Singh, A. Shami, R. Asal, and Y. Li, Software defined networking: State of the art and research challenges, *Computer Networks*, 72: 74-98, 2014.
[22] V. R. Tadinada. Software defined networking: Redefining the future of Internet in IoT and Cloud era, in *Proceedings of the 4th International Conference on Future Internet of Things and Cloud*, Barcelona, Spain, August 22-24, 2014.
[23] Open Networking Foundation, OpenFlow Switch Specification Version 1.5.1, https://www.opennetworking.org/images/stories/downloads/sdn-resources/onf-specifications/openflow/, Accessed on: December 2017.
[24] S. Tomovic, M. Pejanovic-Djurisic, and I. Radusinovic, SDN based mobile networks: Concepts and benefits, *Wireless Personal Communications*, 78(3): 1629-1644 (July 2014).
[25] X. N. Nguyen, D. Saucez, C. Barakat, and T. Turletti, Rules placement problem in OpenFlow networks: A survey, *IEEE Communications Surveys & Tutorials*, 18(2): 1273-1286 (December 2016).
[26] G. Luo, S. Jia, Z. Liu, K. Zhu, and L. Zhang, sdnMAC: A software defined networking based MAC protocol in VANETs, in *Proceedings of the 24th International Symposium on Quality of Service*, Beijing, China, June 20-21, 2016.
[27] Geni, http://groups.geni.net/geni/wiki/OpenFlowDiscoveryProtocol/, Accessed on: 14 March, 2018.
[28] B. Pfaff, J. Pettit, T. Koponen, E. J. Jackson, A. Zhou, J. Rajahalme, J. Gross, A. Wang, J. Stringer, P. Shelar, K. Amidon, and M. Casado. 2015. The design and implementation of open vSwitch, in *Proceedings of the 12th USENIX Conference on Networked Systems Design and Implementation*, Berkeley, CA, USA, May 7-8, 2015.
[29] R. Mijumbi, J. Serrat, J. Rubio-Loyola, N. Bouten, F. De Turck, and S. Latré, Dynamic resource management in SDN-based virtualized networks, in *Proceedings of the 10th International Conference on Network and Service Management*, Rio de Janeiro, Brazil, November 17-21, 2014.
[30] J. Bailey, and S. Stuart, Faucet: Deploying SDN in the enterprise, *ACM Queue*, 14(5): 54-68 (November 2016).

[31] C. Puliafito, E. Mingozzi, and G. Anastasi, Fog computing for the Internet of Mobile Things: Issues and challenges, in *Proceedings of the 3rd International Conference on Smart Computing*, Hong Kong, China, May 29-31, 2017.

[32] A. Mendiola, J. Astorga, E. Jacob, and M. Higuero, A survey on the contributions of Software-Defined Networking to Traffic Engineering, *IEEE Communications Surveys & Tutorials*, 19(2), 918-953 (November 2016).

[33] N. B. Truong, G. M. Lee, and Y. Ghamri-Doudane, Software defined networking-based vehicular ad hoc network with fog computing, in *Proceedings of IFIP/IEEE International Symposium on Integrated Network Management*, Ottawa, ON, Canada, May 11-15, 2015.

[34] K. Bakshi, Considerations for software defined networking (SDN): Approaches and use cases, in *Proceedings of IEEE Aerospace Conference*, Big Sky, MT, USA, March 2-9, 2013.

[35] Open Networking Foundation, SDN Definition, https://www.opennetworking.org/sdn-resources/sdn-definition, Accessed on: November, 2017.

[36] C. J. Bernardos, A. De La Oliva, P. Serrano, A. Banchs, L. M. Contreras, H. Jin, and J. C. Zúñiga, An architecture for software defined wireless networking, *IEEE Wireless Communications*, 21(3), 52-61 (June 2014).

[37] Open Networking Foundation, Northbound Interfaces, https://www.opennetworking.org/images/stories/downloads/working-groups/charter-nbi.pdf, Accessed on: 14 March, 2018.

[38] Open Networking Foundation - Special Report: OpenFlow and SDN - State of the union, https://www.opennetworking.org/images/stories/downloads/sdn-resources/special-reports/Special-Report-OpenFlow-and-SDN-State-of-the-Union-B.pdf, Accessed on: 14 March, 2018.

[39] B. Amento, B. Balasubramanian, R. J. Hall, K. Joshi, G. Jung, and K. H. Purdy, FocusStack: Orchestrating Edge Clouds using location-based focus of attention, in *Proceedings of IEEE/ACM Symposium on Edge Computing*, Washington, DC, USA, October 27-28, 2016.

[40] P. Liu, D. Willis, and S. Banerjee, ParaDrop: Enabling lightweight multi-tenancy at the network's extreme edge, in *Proceedings of IEEE/ACM Symposium on Edge Computing*, Washington, DC, USA, October 27-28, 2016.

[41] B. Varghese, N. Wang, J. Li, and D. S. Nikolopoulos, Edge-as-a-Service: Towards distributed Cloud architectures, in *Proceedings of the 46th International Conference on Parallel Computing*, Bristol, United Kingdom, August 14-17, 2017.

[42] S. Nastic, H. L. Truong, and S. Dustdar, A middleware infrastructure for utility-based provisioning of IoT Cloud systems, in *Proceedings of IEEE/ACM Symposium on Edge Computing*, Washington, DC, USA, October 27-28, 2016.

[43] V. K. Vavilapalli, A. C. Murthy, C. Douglas, S. Agarwal, M. Konar, R. Evans, T. Graves, J. Lowe, H. Shah, S. Seth, B. Saha, C. Curino, O. O'Malley, S. Radia, B. Reed, and E. Baldeschwieler, Apache Hadoop YARN: Yet another resource negotiator, in *Proceedings of the 4th Annual Symposium on Cloud Computing*, Santa Clara, California, October 01-03, 2013.

[44] B. Hindman, A. Konwinski, M. Zaharia, A. Ghodsi, A. D. Joseph, R. Katz, S. Shenker, and I. Stoica, Mesos: A platform for fine-grained resource sharing in the data center, in *Proceedings of the 8th USENIX Conference on Networked Systems Design and Implementation*, Berkeley, CA, USA, March 30 - April 01, 2011.

[45] C. Clark, K. Fraser, S. Hand, J. G. Hansen, E. Jul, C. Limpach, I. Pratt, and A. Warfield, Live migration of virtual machines, in *Proceedings of the 2nd conference on Symposium on Networked Systems Design & Implementation*, Berkeley, CA, USA, May 02-04, 2005.

[46] S. Wang, R. Urgaonkar, M. Zafer, T. He, K. Chan, and K. K. Leung, Dynamic service migration in mobile Edge-Clouds, *IFIP Networking Conference*, 91(C): 205-228 (September 2015).

[47] F. Callegati, and W. Cerroni, Live migration of virtualized Edge networks: Analytical modelling and performance evaluation, in *Proceedings of the IEEE SDN for Future Networks and Services*, Trento, Italy, November 11-13, 2013.

[48] D. Darsena, G. Gelli, A. Manzalini, F. Melito, and F. Verde, Live migration of virtual machines among Edge networks viaWAN links, in *Proceedings of the 22nd Future Network & Mobile Summit*, Lisbon, Portugal, July 3-5, 2013.

[49] S. Shekhar, and A. Gokhale, Dynamic resource management across Cloud-Edge resources for performance-sensitive applications, in *Proceedings of the 17th IEEE/ACM International Symposium on Cluster, Cloud and Grid Computing*, Madrid, Spain, May 14-17, 2017.

[50] G. D'Angelo, S. Ferretti, and V. Ghini, Modelling the Internet of Things: A simulation perspective, in *Proceedings of the International Conference on High Performance Computing Simulation*, Genoa, Italy, July 17-21, 2017.

[51] G. Kecskemeti, G. Casale, D. N. Jha, J. Lyon, and R. Ranjan, Modelling and simulation challenges in Internet of Things, *IEEE Cloud Computing*, 4(1): 62-69 (January 2017).

[52] X. Foukas, G. Patounas, A. Elmokashfi, and M.K. Marina, Network slicing in 5G: Survey and challenges, *IEEE Communications Magazine*, 55(5): 94-100 (May 2017).

[53] Y.C. Hu, M. Patel, D. Sabella, N. Sprecher, and V. Young, Mobile Edge computing—A key technology towards 5G, *ETSI White Paper*, 11(11):1-16 (September 2015).